\documentclass[prb,twocolumn,amsfonts,amssymb,amsmath,showpacs,
  superscriptaddress,a4paper,floatfix]{revtex4}

\usepackage{graphicx}
\usepackage{dcolumn}
\usepackage{bm}

\newcommand{\nup}[1]{n_{#1 \uparrow}}
\newcommand{\ndown}[1]{n_{#1 \downarrow}}
\newcommand{\la}{\langle}
\newcommand{\ra}{\rangle}

\begin{document}

\title{Quantum information analysis of the phase diagram of the
  half-filled extended Hubbard model}

\author{C.\ Mund}
\affiliation{Fachbereich Physik, Philipps-Universit\"at Marburg,
35032 Marburg, Germany}
\author{\"O.\ Legeza}
\affiliation{Research Institute for Solid State Physics and Optics, H-1525
Budapest, P.\ O.\ Box 49, Hungary
}
\author{R.\ M.\ Noack}
\affiliation{Fachbereich Physik, Philipps-Universit\"at Marburg,
35032 Marburg, Germany}

\date{\today}
%
%
%
%
%

\begin{abstract}
We examine the phase diagram of the half-filled one-dimensional
extended Hubbard model using quantum information entropies within the
density-matrix renormalization group.
It is well known that there is a charge-density-wave
phase at large nearest-neighbor and small on-site Coloumb
repulsion and a spin-density-wave at small nearest-neighbor
and large on-site Coloumb repulsion. 
At intermediate Coulomb interaction strength,
we find an additional narrow region of a bond-order phase between these
two phases.
The phase transition line for the transition out of the
charge-density-wave phase changes from first-order at strong coupling 
to second-order in a parameter regime where all three phases are present.
We present evidence that the additional phase-transition line between
the spin-density-wave and bond-order phases is infinite order.
While these results are in agreement with recent numerical work,
our study provides an independent, unbiased means of determining the
phase boundaries by using quantum information analysis, yields values
for the location of some of the phase boundaries that differ from
those previously found, and provides
insight into the limitations of numerical methods in determining phase
boundaries, especially those of infinite-order transitions.
\end{abstract}

\maketitle

\section{Introduction}

The one-dimensional Hubbard model is one of the earliest- and
most-studied models of strongly correlated itinerant
electrons.\cite{hubbard_model}
Its phase diagram at half filling was one of the first treated using
quantum Monte Carlo for quantum lattice systems,\cite{hirsch_qmc_ehm} one of
the first modern numerical methods for these systems. 
Thus it is perhaps surprising that the details of the phase diagram of
this model at half-filling have only recently been established.
\cite{sandvik_tuv,jeckelmann_tuv,satoshis_tuv}

In this work, we determine the ground-state phase boundaries for the
half-filled system using methods based on quantum information, 
\cite{zanardy_entropy,gu2_entropy,vidal_entropy,vidal_entropy_02,
  yang_entropy,deng_block_entropy,ors_2site_entropy}
calculated using the density-matrix renormalization group
(DMRG). \cite{white_dmrg_1,white_dmrg_2,noack_dmrg}
These methods have a number of advantages over other methods involving
gaps, correlation functions, or order parameters.
\cite{ors_2site_entropy,ors_entropy_profile}
First, they involve only the properties of the ground state (or, in
practice, a numerical approximation to the true ground state), and so
can be calculated easily and accurately.
Second, the behavior of the various quantum information entropies can
be directly related to quantum critical behavior, so that they provide
unbiased indicators of quantum phase 
transitions.\cite{gu2_entropy,vidal_entropy,zanardy_entropy,vidal_entropy_02}
Finally, they are quantities both intimately related to the
fundamental approximation of the DMRG and easily and naturally
calculated within the DMRG algorithm. \cite{ors_dbss,ors_qdc}
Comparison with results from other indicators of quantum phase
transitions lends insight into the accuracy and limitations of both
the indicators and the underlying numerical method.

The one dimensional extended Hubbard model\cite{giamarchi_book} (EHM)
is a relatively simple 
model which describes hopping of spin-$\frac{1}{2}$ fermions between
neighboring sites on a lattice. The fermions interact with each
other through on-site and nearest-neighbor interactions, as described
by the Hamiltonian 
\begin{eqnarray}
H = & - &t \sum_{i=1, \sigma}^{N-1} (c^{\dagger}_{i \sigma} c_{i+1 \sigma} +
c^{\dagger}_{i+1 \sigma} c_{i \sigma}) \nonumber\\
& + & U \sum_{i=1}^N n_{i \uparrow} n_{i \downarrow} + V \sum_{i=1}^{N-1} n_{i}
n_{i+1} \, .
\label{eqn:EHM}
\end{eqnarray}
Here $c^{\dagger}_{i \sigma}$($c_{i \sigma}$) creates (annihilates) an electron
with spin $\sigma$ on site $i$, $n_{i \sigma} = c^{\dagger}_{i \sigma} c_{i \sigma}$
and $n_i = n_{i \uparrow} + n_{i \downarrow}$. 
The parameter $t$ denotes the strength of the inter-site hopping, $U$ the
strength of the on-site, and $V$ the strength of the nearest-neighbor
interaction. 
Since the nearest-neighbor and on-site interaction originate
from the (repulsive) Coulomb potential in most materials,
we consider only $U,V > 0$ in this work. 
In general, we will consider finite lattices of size $N$ with open
boundary conditions. 

Although the model is relatively simple, the phase diagram of the
half-filled system is more complicated than one might expect.
When $V=0$, the EHM reduces to the one-dimensional Hubbard
model which is exactly solvable using the Bethe
ansatz.\cite{bethe_solution_hubbard}
From this exact solution, it is known that the system is in a
spin-density-wave (SDW) phase for all $U>0$.
This picture is reinforced by the effective model for strong coupling
($U/t \gg 1$), 
the Heisenberg model, which has a SDW ground state for
antiferromagnetic spin exchange.
One expects the SDW phase to remain the ground state for finite $V$,
as long as $U/V \gg 1$. 
In the opposite limit, $V/U \gg 1$,
one expects the system to form a commensurate charge-density-wave
(CDW) with an alternating spontaneously broken symmetry, i.e., 
sites with average occupation $\langle n_i\rangle > 1$ alternate with
sites with $\langle n_i\rangle < 1$.
Such a picture is obtained both in strong-coupling perturbation
theory\cite{tuv_strong_coupling} and in a weak-coupling renormalization group
treatment ($g$-ology).\cite{tuv_gology}
In fact, the phase boundary between these two phases is predicted to be
at $U=2V$, to lowest order, by both treatments.

While a phase diagram containing only these two phases would not be
particularly interesting in and of itself, a prediction that a
bond-order-wave phase (BOW), a phase characterized by a spontaneously
broken symmetry in which strong and weak bonds alternate, 
occurs near the CDW-SDW phase 
boundary at (approximately) $U=2V$ for intermediate
coupling\cite{nakamura} has sparked new interest in the phase diagram.
In addition, $g$-ology predicts the CDW-SDW phase transition
to be continuous, while strong-coupling theory predicts it to be
first-order, i.e., discontinuous. 
As a result, a number of papers on this issue have been published 
in the last few years, some containing
contradictory
results.\cite{old_sandvik,jeckelmann_tuv,sandvik_tuv,satoshis_tuv}
Apparent agreement in the numerically determined ground-state phase
diagram is present in two 
recent treatments.\cite{sandvik_tuv,satoshis_tuv} 
However, in both of these calculations, the same
quantities, the spin and charge exponents, the bond order parameter,
and the spin and charge gaps,
are used to determine the phase boundaries, and in the latter one
consistency checks or realistic error estimates are not present.
In view of the fact that both underlying numerical methods, Quantum
Monte Carlo (QMC) and the DMRG yield results that are
numerically exact for this system, it is not surprising that the
subsequent similar analyses of the data yields essentially the same results.
As we shall see, alternative ways of determining the phase boundaries
yield more deviation than is indicated by this agreement.

In recent years, various types of quantum entropies have proven
themselves to be a useful and accurate tool for determining the nature
and critical parameter values of quantum phase transitions
(QPTs).
\cite{wootters_2site_entropy,
  vidal_entropy,vidal_entropy_02,yang_entropy,
  wu_block_entropy,gu2_entropy,  ors_2site_entropy,
  molina_2site_entropy, deng_block_entropy, ors_entropy_profile} 
Our goal is to combine these newly developed methods with accurate
numerical calculations to independently determine the phase diagram.

In this article, we calculate the one-site,
two-site, and block entropies of the EHM using the DMRG and use these
quantities to determine the nature and location of the phase
boundaries in its phase diagram.
The remainder of this article is organized as follows:
The concepts and methods used in our calculation will be discussed
in Sec.~\ref{sec:conception}, with Sec.~\ref{sec:quantum_entropy}
defining the one-site, two-site, and block entropies
and Sec.~\ref{sec:dmrg} describing the DMRG method used.
Results for these quantities, calculated using the DMRG, will be
presented in Sec.~\ref{sec:results}.
Sec.~\ref{sec:bond_order_parameter} 
explores the limitations of using quantities such as order parameters
to determine phase boundaries.
Finally, the implications of our results as well as a comparison 
with previous work is contained in Sec.~\ref{sec:conclusion}.

\section{Concepts and Methods}\label{sec:conception}

\subsection{Quantum Entropy}\label{sec:quantum_entropy}

Wu \emph{et al.}~\cite{wu_block_entropy} have argued that, quite
generally, QPTs are 
signalled by a discontinuity in some measure of entanglement in the quantum
system. 
One such measure is the concurrence, \cite{wootters_2site_entropy}
which has been utilized by a number of
authors
\cite{osborne,osterloh,sylju,gu,vidal_entropy,vidal_entropy_02,
  roscilde,yang_entropy} 
in their studies of spin models. 
The local measure of entanglement, the 
one-site entropy, \cite{preskill_quantum_information}
has been proposed by
Zanardi \cite{zanardy_entropy} and Gu \emph{et al.} \cite{gu2_entropy},
or the negativity, 
by Vidal \emph{et al.} \cite{vidal_2site_entropy_negativity} 
to identify QPTs.
All of these quantities exhibit anomalies, i.e., either discontinuities in the
quantity itself or in its
higher derivatives or extrema,
at points in parameter space that correspond to QPTs in exactly
solvable models.
What kind of anomaly occurs depends on the quantity and on the type of
QPT; sometimes the anomaly appears only in some but not all of the quantum
information entropies.\cite{ors_2site_entropy} 
However, it has become clear from recent work
\cite{deng_block_entropy,ors_entropy_profile} that, when the
appropriate variant of the quantum information entropy is chosen, it
can be used as a particularly convenient and accurate probe of the
QPT.
One example of this is the two-site entropy, a good probe of
transitions to dimerized phases such as the BOW phase we study 
here.\cite{ors_2site_entropy} 

All of the quantum entropies discussed here have as their starting
point the reduced density matrix of a bipartite system.
Assuming that a quantum mechanical system in a pure state $|\Psi \ra$ is
divided into two parts,  part A with complete orthonormal basis $|i \ra$ and 
part B with complete orthonormal basis $|j \ra$, the wave function can
be expressed as 
\begin{equation}
|\Psi \ra = \sum_{i,j} C_{ij} \, |i \ra \otimes |j \ra \, .
\end{equation}
The density matrix of the entire system is defined as
\begin{equation}
\rho = |\Psi \ra \la \Psi | \, ,
\end{equation}
and the reduced density matrix for subsystem A is
\begin{subequations}
\label{eqn:reduced_den_mat}
\begin{eqnarray}
\rho_{A} & = & \mbox{tr}_B \left( |\Psi \ra \la \Psi | \right) \\
&  = & \sum_j
\la j | \Psi \ra \la \Psi | j \ra \label{eqn:reduced_den_mat_b}\\
& = & \sum_{i, i^\prime, j}
 C_{ij}  C^*_{i^\prime j} |i \ra\la i^\prime|
 \label{eqn:reduced_den_mat_c} \, .
\end{eqnarray}
\end{subequations}
The von Neumann entropy for subsystem A is then 
\begin{equation}\label{eqn:von_neumann_entropy}
  S_A = - \text{Tr}\;  \rho_A \, \ln \rho_A = -\sum_\alpha \rho_\alpha
  \ln (\rho_\alpha) \, , 
\end{equation}
where $\rho_\alpha$ are the eigenvalues of the density matrix $\rho_A$.

It is also useful to define the density matrix in terms of an
(unnormalized) projector which takes subsystem A
from state $|i^\prime\ra$ to state $|i\ra$, 
\begin{equation}
  \la i |  \rho_{A} | i^\prime  \ra 
  \equiv \la \Psi | P_{i^\prime, i} |  \Psi \ra
  = \la \Psi | \left ( \sum_j  
  | i \ra \otimes | j \ra
  \la j | \otimes \la i' |  \right ) | \Psi \ra \, .
\label{eqn:den_mat_projector}
\end{equation}
For simple subsystems and an appropriate choice of basis $|i\ra$, 
the operator $P_{i^\prime,i}$ can be expressed in terms of
relatively simple operators, so that its matrix elements can be
directly calculated using the corresponding
observables;\cite{rissler_quantum_entropy} this will be used in the
following.

\subsubsection{One-site entropy}

In order to form the one-site entropy, which we will denote
$S_\ell(1)$, the subsystem A is simply
taken to be a particular single site $\ell$.
Since Hamiltonian\ (\ref{eqn:EHM}) contains no spin-flip processes, 
the reduced density matrix can be obtained directly in
diagonal form if we take the spin occupation basis $\left( |0\rangle, | \uparrow
\rangle, | \downarrow \rangle, | \uparrow \downarrow \rangle \right)$
as the basis states $|i\ra$ in Eq.~(\ref{eqn:den_mat_projector}).
In this basis, we can directly obtain the eigenvalues from the
expectation values of the four operators \cite{rissler_quantum_entropy}
\begin{eqnarray*}
\la \uparrow \downarrow |\rho_\ell| \uparrow\downarrow\ra 
& = & \la \nup{\ell} \ndown{\ell} \ra \\
\la \downarrow |\rho_\ell| \downarrow\ra 
& = & \la (1- \nup{\ell}) \ndown{\ell} \ra \\
\la \uparrow |\rho_\ell| \uparrow\ra 
& = & \la \nup{\ell} (1- \ndown{\ell}) \ra \\
\la 0 |\rho_\ell| 0 \ra 
& = & \la (1- \nup{\ell})(1- \ndown{\ell}) \ra \, .
\end{eqnarray*}
The one-site entropy is relatively easy to calculate because it requires
only four local measurements on site $\ell$. 
While the one-site entropy is useful in some cases for characterizing
first-order QPTs, 
it is typically not well suited  
for determining higher order QPTs because
anomalies are often only discernible for large system sizes and are sometimes
not present at all.\cite{gu2_entropy,ors_2site_entropy} 
In particular, changes in intersite bond strength have no
influence on it. 
Since the one-site entropy necessarily only depends on quantities that
are localized on that one site (in fact, only on the average
occupancy and on the average double occupancy), it cannot contain
spatial information that is nonlocal.

\subsubsection{Two-site entropy}
\label{sec:2_site_entropy}

It is therefore often useful to examine the von Neumann entropy associated 
with a larger subsystem.
One convenient choice of subsystem A is that of two sites $p$ and $q$, 
i.e., the two-site entropy $S_{p,q}$.
In particular, we are interested in characterizing a BOW phase, which, for
open boundary conditions, will have a broken bond-centered
spatial symmetry.
Thus, we are principally interested in the behavior of the two-site
entropy for different bonds, i.e., we take $p$ and $q$ to be pairs of
nearest-neighbor sites.
Therefore the two-site entropy will be denoted as $S_p(2)$ in this paper,
meaning $S_{p,p+1}$.

The two-site entropy can be obtained by calculating and diagonalizing
the reduced density matrix for the two sites.
As for the one-site
density matrix, its matrix elements in the occupation number basis can
be expressed straightforwardly in terms of expectation values
localized to the two sites by considering the projector of
Eq.~(\ref{eqn:den_mat_projector}).
However, the result is necessarily somewhat more complicated than for the
one-site density matrix.
Since spin and particle number are conserved quantum
numbers in Hamiltonian\ (\ref{eqn:EHM}), the resulting reduced density
matrix is block diagonal (rather than diagonal as for the one-site
case) and has 26 independent
matrix elements; for details, see
Ref.~\onlinecite{rissler_quantum_entropy}.
Hence, 26 independent measurements, followed by the appropriate
matrix diagonalization and the summation of
Eq.~\eqref{eqn:von_neumann_entropy}, must be performed for every
two-site entropy calculated. 

Note that calculating all $N (N-1)$ two-site entropies would be
prohibitively expensive for large system sizes. 
While considering only nearest-neighbor bonds reduces this to $N-1$,
it is usually sufficient to calculate the
entropy of the two pairs of innermost sites, $S_{N/2-1}(2)$ and 
$S_{N/2}(2)$, i.e., two bonds chosen to be in the middle of the
system to minimize boundary effects, to characterize a BOW phase.

\subsubsection{Block entropy}
\label{sec:block_entropy}

The block entropy is also based on splitting the system into two subsystems
$A$ and $B$.
However, subsystem $A$ is now taken to contain the $\ell$ contiguous
sites $1$ to $\ell$ and subsystem $B$ to contain the remainder, sites
$\ell +1$ to $N$.
(Note that other choices of sets of contiguous sites are also possible.)
We then calculate the von Neumann entropy $S(\ell)$ using 
Eq.~\eqref{eqn:von_neumann_entropy} for $\ell \in {1,...,N}$. 
Note that calculating the projector $P_{i^\prime,i}$ of 
Eq.~\eqref{eqn:den_mat_projector} would be prohibitively complicated
and expensive. 
Here, we instead calculate the density matrix
directly from the wave function using
Eq.~\eqref{eqn:reduced_den_mat_c}.
Fortunately, the wave function for all divisions of the system $\ell$
is readily available in the appropriate finite-system step of the DMRG
algorithm, and, in fact, the corresponding von Neumann entropy 
is intimately related to the approximation made in the 
DMRG.\cite{ors_dbss,ors_qdc}

Unlike the two-site entropy, which has an finite upper bound ($\ln 4$ for
Hubbard-like models), the block
entropy generally grows as $\mathcal{O} \left( \ln N \right)$ for
critical one-dimensional systems \cite{vidal_block_entropy} (but
scales to a finite value for non-critical systems).
Although such a potential divergence would at first glance seem favorable
for  studying QPTs, the situation is actually more complicated:
boundary effects from open boundary conditions should have a stronger
influence on the block entropy than on a more locally defined quantity
such as the one-site or two-site entropy.
It is therefore not clear which of these entropies can more
accurately detect QPTs, but it seems sensible to expect that
fast growing peaks are better detected by the block entropy, while 
non-diverging anomalies, such as discontinuities in derivatives,
can be more precisely determined by the two-site entropy.

\subsection{DMRG}\label{sec:dmrg}

The DMRG calculations were carried out using the finite-system 
algorithm\cite{white_dmrg_1,white_dmrg_2,noack_dmrg} on 
systems with (mostly) open boundary conditions with from $N=32$ to
$N=512$ lattice sites.
Open rather than periodic boundary conditions were used for two
reasons: First, the DMRG is substantially more accurate for fixed system
size and computational effort.
Second, since open boundary conditions explicitly break the
translational invariance, a corresponding spontaneously broken
symmetry of the ground state, which, strictly speaking, can only occur in the
thermodynamic limit, appears 
in the entropy profiles of finite-sized systems.
\cite{laflorence_entropy,ors_entropy_profile}
This is, for example, the case for the BOW phase.

Since only ground-state properties are required to calculate the von
Neumann entropies, we need only calculate the ground-state wave
function and the appropriate observables for the half-filled system.
We have used dynamic block-state selection (DBSS), which chooses the
size of the Hilbert space retained in each truncation by keeping the
block entropy of the discarded density-matrix eigenstates 
constant.\cite{ors_qdc}
It is important to do this because the accuracy of the von Neumann
entropy as calculated using the approximate DMRG  wave function in
Eq.~\eqref{eqn:reduced_den_mat_c} is directly related to the entropy
threshold used.
In our calculations we have used a threshold for the quantum information
loss\cite{ors_qdc} of $\chi < 10^{-10}$, which yields extremely
precise results and requires a maximum of approximately $m=3000$ block
states to be retained.
In general, we estimate that the errors due to the truncation in
the DMRG calculations are negligible in comparison to the uncertainties
arising from the finite-size scaling.

\subsection{Limitations of other quantities}
\label{sec:quantity_limitations}

Naively, the most straightforward way to 
determine the existence and extent of a BOW phase would
probably be to examine the bond order parameter, defined as
\begin{equation}
  B = \dfrac{1}{N} \sum_{i=1,\sigma}^{N-1} \langle 
  c_{i+1 \sigma}^\dagger c_{i \sigma} + c_{i \sigma}^\dagger c_{i+1 \sigma} \rangle \, .
\end{equation}
Whether the bond order parameter is finite or vanishing in the
thermodynamic limit would determine whether a given point in parameter space is
ordered or not, and a grid of such points can be used to determine the
phase boundaries.

Unfortunately, there are two major problems with this strategy. 
First, the transition between the BOW and SDW phases is expected to be
infinite-order, and we indeed
find behavior characteristic of an infinite-order transition.
This means that the extrapolated bond order parameter tends to zero
exponentially as the BOW-to-SDW transition is approached.
Second, while it is know that the bond order parameter is linear in
$1/N$ in the CDW 
phase, where $N$ is the system size, and proportional to $1/\sqrt{N}$
in the SDW phase, 
the analytic form of the finite-size scaling in the BOW phase is not
known and changes nature as the transition is approached.
We will examine this issue in more detail in 
Sec.\ \ref{sec:bond_order_parameter}.

Unfortunately, similar difficulties arise in other quantities that are
typically used to determine critical parameters. 
In particular, the charge gap goes to zero in a clear manner at the
CDW-BOW phase transition, but does not 
exhibit any anomalies at the BOW-SDW transition.
The spin gap, on the other hand, does go to zero at the BOW-SDW transition,
but, like the bond order parameter, with an exponential dependence.
Therefore, it is also not well-suited for exactly determining the
phase boundary.

\section{Results}\label{sec:results}

\begin{figure}[bth]
\includegraphics[angle=270,width=0.48\textwidth]{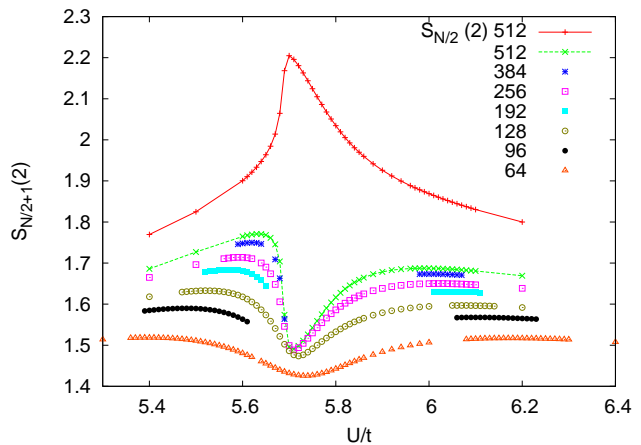}
\caption{(Color online) \label{fig:2sV30} Two-site entropy $S_{N/2+1}(2)$ for
  $V/t=3$, plotted as a function of $U/t$. 
  System sizes range from $N=64$ to $N=512$ sites.
}
\end{figure}

\begin{figure}[bth]
\includegraphics[angle=270,width=0.48\textwidth]{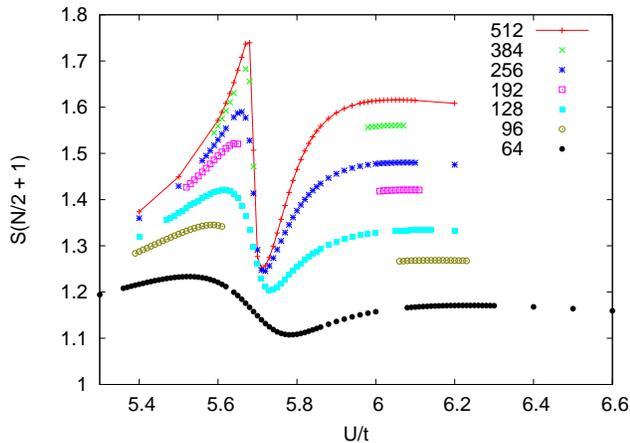}
\caption{(Color online) \label{fig:BlockV30} Block entropy at the
  center of the chain $S(N/2+1)$ for $V/t=3$ and system
  sizes from $N=64$ to $N=512$ sites.
} 
\end{figure}

We consider first the behavior of the two-site and block entropies.
In order to map out the phase boundaries, we have swept $V$ from $V=2$ to
$V=5.5$ in steps of $0.5$ for a number of values of $U$.
For all sweeps, we find that two peaks develop in both the two-site
entropy, Fig.~\ref{fig:2sV30}, and in the block
entropy, Fig.~\ref{fig:BlockV30}, for sufficiently large systems.
Note that the second peak in the block entropy can only be seen at
system sizes of $N=96$ and larger.
This slow size dependence might be the reason why this phase was not seen
by Deng {\sl et. al.} \cite{deng_block_entropy}.

We interpret the peak at lower $U$ as marking the CDW-BOW phase
transition at the corresponding system size, and the peak at larger $U$
as indicating the BOW-SDW phase transition. 
The differing shapes of the two peaks are consistent with the picture that the
CDW-BOW transition is first order in this parameter regime, \cite{sandvik_tuv}
while the BOW-SDW transition shows characteristics of an
infinite-order transition.

The interpretation that the intervening phase is a BOW phase is
supported by the behavior of the two-site entropy for two adjacent
(odd and even) bonds in the center of the lattice, plotted for the
largest system size, $N=512$, in Fig.\ \ref{fig:2sV30}.
The dimerization entropy is given by $D_s=S_{N/2+1}(2)-S_{N/2}(2)$,
which is the difference between the two $N=512$ curves.
It is clear that this difference reaches a marked maximum between the
two peaks associated with the phase transitions.
Finite-size extrapolation (not shown) indicates that $D_s$ remains
finite in the thermodynamic limit in the intermediate phase.

\begin{figure}[bth]
\includegraphics[angle=270,width=0.45\textwidth]{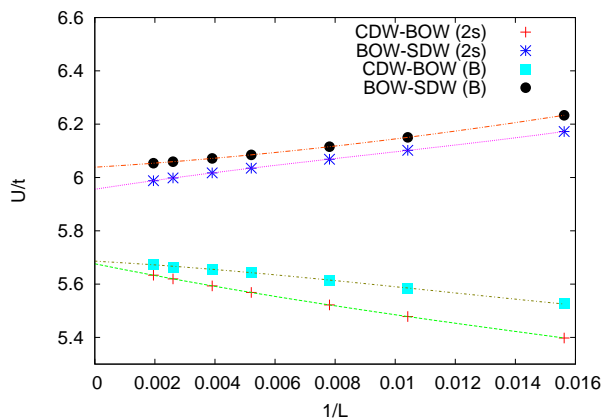}
\caption{(Color online) \label{fig:EntFitv30} Finite-size
  extrapolation of the peaks in 
  Figs.\ \ref{fig:2sV30} and \ref{fig:BlockV30} to the thermodynamic
  limit, $1/L\to 0$, using fourth order polynomials in $1/L$.
  ``2s'' labels the two-site entropy and ``B'' the block entropy.
} 
\end{figure}

The positions of the two peaks in the two-site or block entropy  can
then be extrapolated to the thermodynamic limit. 
While the functional form of this extrapolation is not exactly known,
using a fourth-order polynomial yields stable results, with a rapid
falloff in coefficient size for higher orders. 
Therefore, the behavior is predominantly linear. 
This can be seen in Fig.~\ref{fig:EntFitv30}.

\subsection{The CDW-BOW transition}\label{sec:cdw_bow}

\begin{figure}[bth]
\includegraphics[angle=270,width=0.48\textwidth]{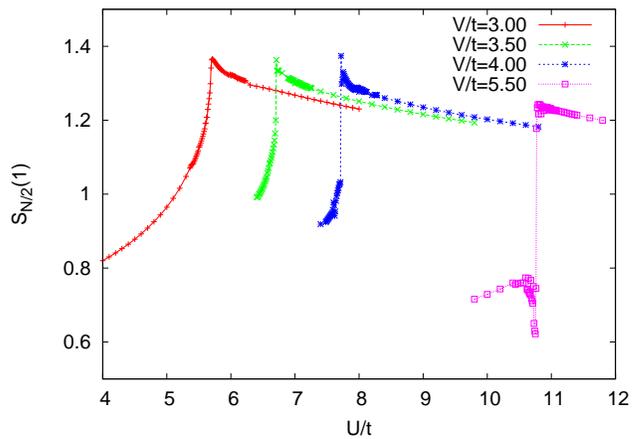}
\caption{(Color online) \label{fig:1sEntFitInfty} Extrapolated
  one-site entropy $S_{N/2}(1)$ 
  plotted as a function of $U/t$ for various values of $V/t$.
}
\end{figure}

A comparison of Figs.~\ref{fig:2sV30} and \ref{fig:BlockV30} clearly shows that
the peaks in the block entropy at lower $U$ are sharper and higher
than those in the two-site entropy. 
In addition, the position of the peaks in the block entropy
match the jump in the one-site entropy (see
Fig.\ \ref{fig:1sEntFitInfty}) better  
than the position of the peak in the two-site entropy. 
Therefore, we conclude that the block entropy is, in general,
a better indicator of the position of the CDW-BOW transition than the
two-site entropy. 
As can be seen in Fig.~\ref{fig:EntFitv30}, the fits to the two-site
and the block entropies (the two lower curves) match almost exactly in
the thermodynamic limit, so that this issue is virtually irrelevant here.

A more difficult issue is to determine where nature of the CDW-BOW
transition changes from first-order to continuous. 
This can best be investigated using the finite-size extrapolation of
the one-site entropy.
In Fig.~\ref{fig:1sEntFitInfty}, we show the one-site
entropy extrapolated to the thermodynamic limit using a polynomial of
cubic order.
It is apparent that the entropy has
a jump at the transition point when $V/t \geq 4$, a clear indication
of a first-order transition. 
At approximately $V/t = 3$, the transition is close to becoming
continuous in that a jump is no longer present.
For smaller values of $V/t$ (not shown), it is clearly continuous. 
Therefore, we conclude
that the first-order-to-continuous bicritical point must occur
somewhere near, but below $V=3$. 
Note that it is difficult to determine the location of this point with
more accuracy because one would have to determine whether or not an
increasingly small jump is
present in the finite-size extrapolated data as the bicritical point
is approached on a sufficiently finite grid.

\subsection{The BOW-SDW transition}\label{sec:bow_sdw}

The BOW-SDW transition is believed to be infinite-order
\cite{giamarchi_book} 
and is therefore much harder to characterize. 
While we were able to obtain a fairly good
estimate for the position of the CDW-BOW transition just by examining
the bond order parameter, this is not possible for the BOW-SDW
transition, as discussed in Sec.~\ref{sec:bond_order_parameter}. 
As can be seen in Figs.\ \ref{fig:2sV30} and \ref{fig:BlockV30}, the
maxima for the BOW-SDW transition (the peaks at higher $U/t$) are much
broader. 
Therefore, small errors in the numerical calculations
would have a bigger influence on the result than for the case of the
CDW-BOW transition.
Due to this, we have carried out very high precision calculations, as
already described in Sec.~\ref{sec:dmrg}.

\begin{figure}[bth]
\includegraphics[angle=270,width=0.48\textwidth]{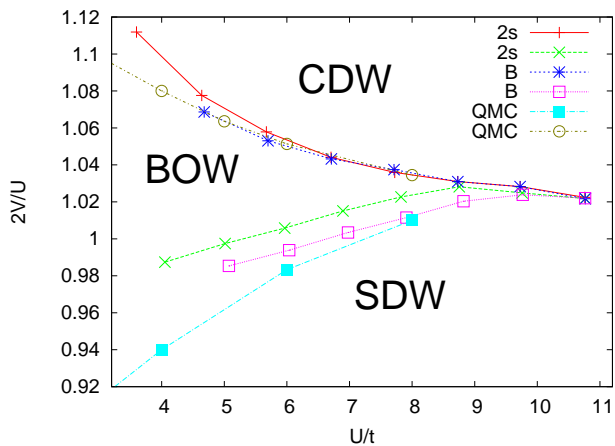}
\caption{(Color online) \label{fig:UVPlaneTilted} The phase diagram in
  the tilted $U$-$V$-plane, showing phase boundaries determined using
  the two-site entropy (2s), the block entropy (B), and including
  results from Ref.\ \onlinecite{sandvik_tuv}, determined from QMC
  calculations. 
} 
\end{figure}

Results for the phase boundaries are shown in
Fig.~\ref{fig:UVPlaneTilted}, plotted in the tilted $V$-$U$ phase,
i.e., with axes $2V/U$ and $U$, so that the transition region is
discernible. 
Included are data from
Ref.\ \onlinecite{sandvik_tuv}, in which the phase boundaries were
determined from the spin and charge exponents calculated using QMC
methods.
There is generally very good agreement for the location of the CDW-BOW
transition, with some deviation of the results from the two-site
entropy at smaller $U/t$ values.
As we have argued above, the block entropy is a better indicator of
the position of this transition because the peak is better developed
and grows more rapidly with system size.

For the BOW-SDW transition, the
two-site and the block entropies coincide perfectly upon scaling for
higher $U$ and $V$ values, where only a first-order transition is
present, as can be seen in Fig.~\ref{fig:UVPlaneTilted}.
However, there is a discrepancy
in the position of the peaks in the infinite-system extrapolations of
the two-site and the block entropies (see Fig.~\ref{fig:EntFitv30})
for smaller values of $U$ and $V$. 
This occurs for the entire range of parameter values in which a BOW
phase and, therefore, an infinite-order transition is present.
This is probably partially due to uncertainty in localizing the broad
peak in the entropies corresponding to the BOW-SDW transition.
In addition, due to the strong increase of the
block entropy  with system size below $U/t\approx 4$, we cannot treat systems
of more than a few hundred lattice sites for fixed $\chi=10^{-10}$,
leading to increased uncertainty in the finite-size extrapolation.
For $U/t<3$ the second peak in the entropy
functions develop only for $N\ge 96$, quite severely limiting the
extrapolation to infinite system size.
We therefore display the phase diagram only for $U/t>3$ in
Fig.\ \ref{fig:UVPlaneTilted}

\subsection{Bond-order-parameter results}
\label{sec:bond_order_parameter}

\begin{figure}[bth]
\includegraphics[angle=270,width=0.48\textwidth]{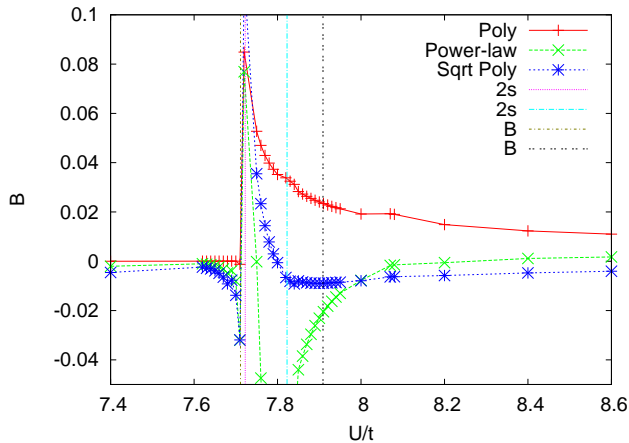}
\caption{(Color online) \label{fig:BOparameter} 
  Bond order parameter $B$ extrapolated to the thermodynamic limit using
  fits to three different functions: fit to a third-order polynomial
  in $1/N$ (Poly), fit to a third-order polynomial in $1/\sqrt{N}$
  (Sqrt-Poly), and fit to a power law $1/N^\alpha$ (Power-law) with
  $\alpha$ a fitting 
  parameter, plotted as a function of $U/t$.
  The vertical lines indicate the transition point determined using
  the two-site entropy (2s) and the block entropy (B).
}
\end{figure}

As already mentioned in Sec.~\ref{sec:quantity_limitations}, we have
carried out calculations for the bond order parameter to very high
precision.
The resulting data are sufficiently accurate so that they
can be regarded as essentially exact for a particular size for fitting
purposes.
We have carried out the extrapolation to the infinite-system limit
by fitting to three different functions: a polynomial in $1/N$, a
polynomial in $1/\sqrt{N}$, and a power law of the form
$1/N^{\alpha}$.
The extrapolated data are shown in Fig.~\ref{fig:BOparameter}.
In the CDW phase, to the left of the transition indicated by vertical
lines,
the fit to a polynomial in $1/N$ gives the best result, yielding the
expected value, zero, to the best accuracy.
In the SDW phase, for $U/t \gtrsim 8$,
the polynomial fit in $1/\sqrt{N}$ and the fit to a power law work
better, yielding the expected value of zero to within reasonable
accuracy.
This is indicative of a scaling whose dominant term falls off more
slowly than $1/N$.
In the intermediate region, i.e., in the BOW phase, the results differ
significantly, with both the fit to powers of $1/\sqrt{N}$ and the
power-law fit extrapolating to spurious negative values. 
In addition, the power-law fit is clearly unstable in the BOW region.
While the fit to a polynomial in $1/N$ seems to be more stable, it
clearly overestimates the bond order parameter significantly in both
the SDW phase and in most of the BOW region.

Therefore, we conclude that the behavior of the bond order parameter
can be used to confirm the position of the CDW-BOW phase transition
but is notoriously unreliable for 
determining the location of the BOW-SDW phase
transition.

\section{Phase Diagram and Discussion}\label{sec:conclusion}

\begin{figure}[bth]
\includegraphics[angle=270,width=0.48\textwidth]{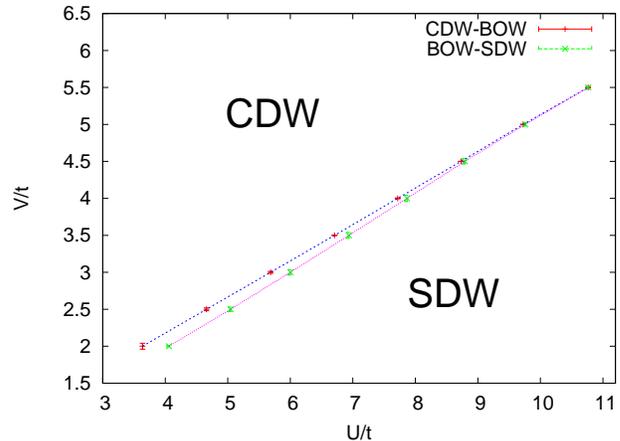}
\caption{(Color online) \label{fig:UVPlane} The phase diagram in the U-V-Plane
  obtained from our calculations.
  On this scale, uncertainties in the position of the transition lines
  are smaller or about the size of the symbols.
  The phase in the narrow region between the two transition lines is
  BOW.
  }
\end{figure}

We now summarize the current state of knowledge of the phase diagram
of the half-filled extended Hubbard model and discuss open issues and
uncertainties.
The overall phase diagram is relatively well understood; our results
are depicted in Fig.\ \ref{fig:UVPlane}.
The presence of the SDW phase at $V \ll 2 U$ and the CDW phase at 
$ V\gg 2U$ have been long understood, as well as the fact that the
transition occurs at $V \approx 2U$.
The picture of there being a single first-order transition line at
strong $U$ and $V$ \cite{tuv_strong_coupling} is also
well-established.
Our work lends support to a picture in which an intermediate BOW
phase is present between the CDW and SDW phases for intermediate to
small $U$ and $V$; our results indicate that this phase is present for
$V/t \lesssim 5$.
At this point, we find that the first-order CDW-SDW transition line
bifurcates into a first-order CDW-BOW transition line and an infinite-order
BOW-SDW transition line.
The CDW-BOW transition line remains first-order at a bicritical point
at somewhat smaller $V/t$, below which it becomes continuous, presumably
second-order.
These results are in reasonable agreement with the results of
Refs.\ \onlinecite{sandvik_tuv} and \onlinecite{satoshis_tuv}.
We therefore regard these features of the phase diagram as being
well-established.

We now discuss details of the phase transition more quantitatively.
As we have seen in Fig.~\ref{fig:UVPlaneTilted}, there is not much
uncertainty in the position of the CDW-BOW phase transition. 
The remaining interesting question
for this transition is the location of the bicritical point,
i.e., exactly where the phase-transition line goes from being
first-order to being continuous (presumably second-order). 
However, as we have pointed out in Sec.~\ref{sec:cdw_bow},
entropy measurements can only roughly determine that this point
occurs at around $V/t=3$ and are not an ideal measurement to locate
it more accurately.
While other authors have obtained putatively more accurate values for
the location of this bicritical point, \cite{satoshis_tuv,sandvik_tuv}
we point out that the inaccuracies in our method reflect intrinsic
limitations of the numerical methods, which stem both from the DMRG
truncation error as well as from the limitations of working with
finite systems.

The exact position of the BOW-SDW phase-transition line is also
somewhat uncertain.
There is a small, but significant, discrepancy in our calculations
between the values obtained from the two-site-entropy and those
obtained from the block-entropy; however,
both of these extrapolated values seem to converge smoothly to
the same line at the tricritical point.
The most likely explanation for the discrepancy in the extrapolations
lies in the finite-size extrapolation, i.e., 
more precise results could be obtained if larger system sizes could be
treated. 
The deviations between the two values for the transition line can be
taken as a rough estimate of the uncertainty in the position of the
line.
In addition, both of these values deviate from
those of Refs.\ \onlinecite{sandvik_tuv} and \onlinecite{satoshis_tuv}
(see Fig.~\ref{fig:UVPlaneTilted}), a deviation to larger values of
$V$, i.e., to a narrower BOW phase, in both cases.

In summary, the DMRG method, coupled with the use of single-site,
two-site, and block entropies, is a powerful, relatively unbiased
method to determine subtle properties of phase diagrams such as that
of the extended Hubbard model at half filling.
However, the limitations of the numerical results reflect the intrinsic
limitations of the method and the problem studied.
In particular, three aspects of the phase diagram studied here remain
difficult to pin down numerically: the exact position at which the
CDW-SDW line bifurcates into CDW-BOW and BOW-CDW transition lines, the
position of the bicritical point at which the CDW-BOW transition goes
from first to second order, and the exact position of the
infinite-order BOW-SDW transition.
In addition, our calculations make clear that quantitative
determination of the transition lines become very difficult in the
region of small $U$ and $V$.

\acknowledgments
The authors thank A.\ Sandvik for providing data from
Ref.\ \onlinecite{sandvik_tuv} in numerical form and F.\ Gebhard for
helpful discussions.
This research was supported in part by the Hungarian Research Fund (OTKA),
Grants Nos.\ K 68340 and K 73455, and by the J\'anos Bolyai Research
Fund.  

\bibliography{biblio}

\end{document}